\documentclass[%
 aip,
 amsmath,amssymb,
preprint,%
]{revtex4-1}

\usepackage{graphicx}
\usepackage{dcolumn}
\usepackage{bm}

\usepackage[utf8]{inputenc}
\usepackage[T1]{fontenc}
\usepackage{mathptmx}
\usepackage{etoolbox}

\makeatletter
\def\@email#1#2{%
 \endgroup
 \patchcmd{\titleblock@produce}
  {\frontmatter@RRAPformat}
  {\frontmatter@RRAPformat{\produce@RRAP{*#1\href{mailto:#2}{#2}}}\frontmatter@RRAPformat}
  {}{}
}%
\makeatother
\begin{document}

\preprint{AIP/123-QED}

\title[RIT-2.5 Radiofrequency Ion Thruster]{Experimental Characterization and Simulation-Based Analysis of the RIT-2.5 Radiofrequency Ion Thruster}
\author{N. Joshi}
\author{O. Meusel}%
 \email{email@ninadjoshi.de.}
\affiliation{ 
Institut für Angewandte Physik, Goethe Universität, Frankfurt am Main, Germany
}%

 \homepage{http://www.ninadjoshi.de.}

\date{\today}

\begin{abstract}
We report on the experimental characterization of the RIT-2.5 radiofrequency ion thruster, complemented by particle-in-cell (PIC) simulations to establish correlations between plasma parameters and extracted beam properties. 
The thruster was operated in a dedicated vacuum facility and equipped with optical spectroscopy, Faraday cup, and retarding field energy analyzer (RFEA) diagnostics. 
Electron temperature was determined from optical emission line ratios ($T_e \approx 3$–$7~\mathrm{eV}$), while the energy distribution of extracted ions was measured using the RFEA. 
A comparative simulation campaign was performed to reproduce the experimentally observed energy spectra and to extract plasma density values otherwise inaccessible due to the lack of intrusive diagnostics, yielding best-fit estimates of $n_e \approx 1.2\times10^{16}~\mathrm{m^{-3}}$. 
The results demonstrate that the energy spread of the extracted beam is strongly dependent on plasma density and electron temperature. 
The combined experimental–numerical approach provides a non-intrusive yet robust pathway for performance optimization of RF ion thrusters, offers validation benchmarks for advanced plasma simulation codes, and is particularly well-suited to compact µN-class thrusters for nanosatellite attitude and orbit control.
\end{abstract}

\maketitle

\section{\label{sec:level1}Introduction
}

Radiofrequency ion thrusters (RITs) have been developed and deployed as efficient electric propulsion systems for space applications for more than four decades~\cite{Kaufman1960, Loeb2002, Loeb}. Their ability to generate high specific impulse with comparatively low propellant consumption has made them a cornerstone technology for long-duration satellite station-keeping and deep-space missions. 

Miniaturized thrusters with diameters of only a few centimeters, such as the RIT-2.5, deliver thrusts on the order of micro-Newtons ($\mu$N-RITs) and are critical for ultra-fine spacecraft positioning and nanosatellite attitude control~\cite{Samples2019,Yang2022,Koizumi2014}. In this size class, precise control over beam current, divergence, and energy spread is essential, since even small performance variations can impact formation flying, drag compensation, or rendezvous maneuvers.

Despite their maturity, predictive modeling and quantitative performance characterization of RITs remain active areas of research. This is largely due to the complexity of radiofrequency plasma discharges, where plasma density, electron temperature, and sheath dynamics jointly determine the extracted ion beam current and divergence~\cite{Okawa2004,Ott}. Accurate measurements of these parameters are crucial both for thruster optimization and for the validation of numerical plasma models, such as particle-in-cell (PIC) or collisional–radiative (CR) approaches. 

Direct probing of the discharge plasma is often restricted by the presence of strong RF fields and limited physical access, particularly in small-scale thrusters. As a result, indirect diagnostic techniques such as optical emission spectroscopy (OES) and retarding field energy analysis (RFEA) have become indispensable. While OES allows for non-intrusive determination of the electron temperature, RFEA provides insight into the longitudinal energy distribution of the extracted ion beam. 

In this work, we address this diagnostic gap by combining OES and RFEA with dedicated PIC simulations to infer otherwise inaccessible plasma parameters, in particular the electron density inside the discharge chamber. We present a comprehensive experimental study of the RIT-2.5, describe the diagnostic setup, discuss the extraction of plasma parameters, and establish a quantitative link between the measured energy spread of the beam and the underlying plasma density. The methodology developed here not only supports the performance optimization of µN-class RITs but also provides benchmark data for the validation and further development of advanced numerical tools.

\section{\label{Geo}Geometry of RIT-2.5}

The RIT-2.5 features a diameter of $2.5~\textrm{cm}$, designed with precision for optimal performance. Its discharge chamber boasts an innovative combination of a dome shape on one side and a cylindrical section on the extraction side. The dome, a half-sphere with a radius of $12.5~\textrm{mm}$, aligns perfectly with the radius of the cylindrical part, ensuring seamless integration. The cylindrical section itself measures $9.9~\textrm{mm}$  in length, embodying efficiency in design.

%
\begin{figure}
\includegraphics[width=80mm]{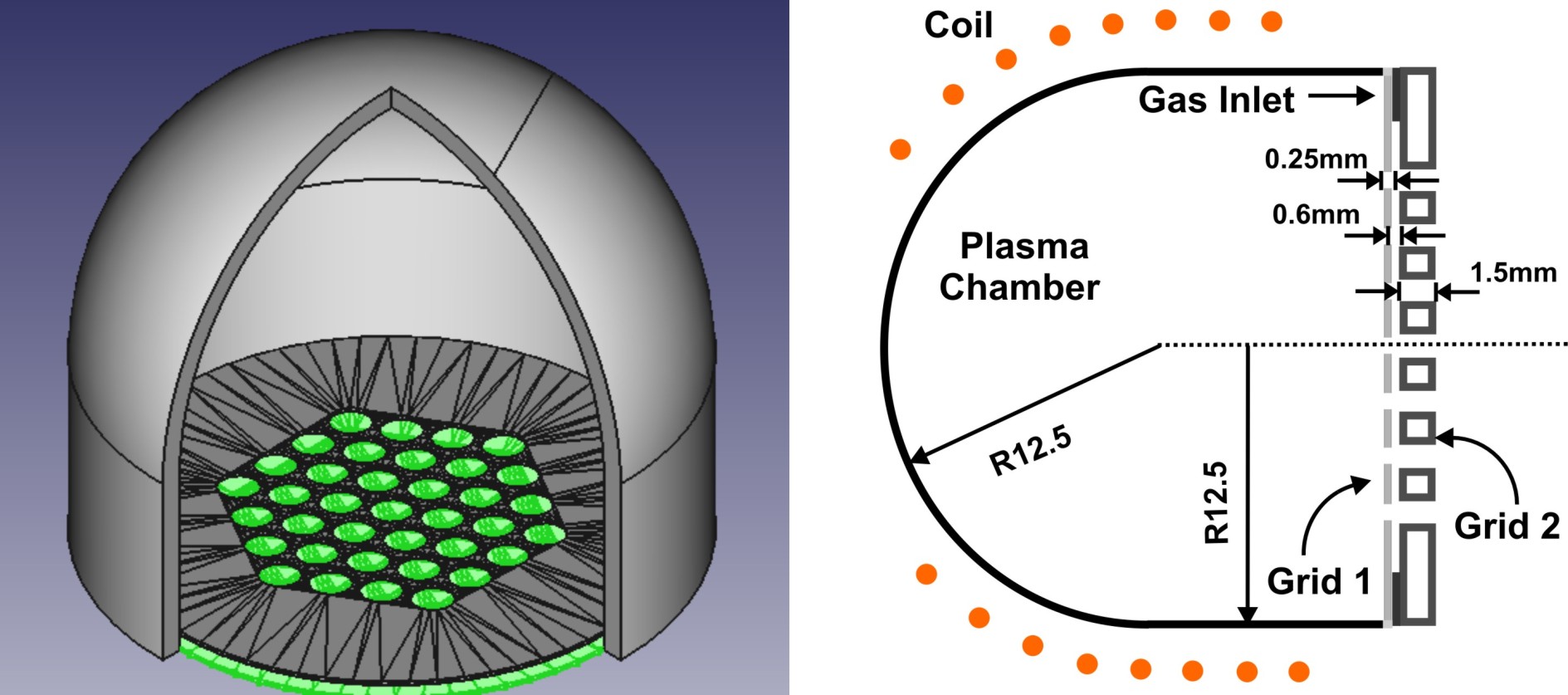}
\caption{\label{fig:geometry} A 3-Dimensional CAD model and a 2-dimensional drawing of the RIT-2.5 ion thruster.}
\end{figure}

To enhance plasma coupling, a sophisticated coil system consists of eight individual coils that surround the plasma chamber with rotational symmetry. This setup, with four coils around the cylindrical portion and four around the dome, maximizes RF power transfer, making the system more effective.

The extraction section is engineered for precision, featuring two grids separated by a gap of $20.85~\textrm{mm}$. The first grid, with a thickness of $0.25~\textrm{mm}$, and the second grid, measuring $1.5~\textrm{mm}$, host 37 extraction holes arranged in an elegant six-fold symmetry. The first grid's holes have a radius of $0.95~\textrm{mm}$, while the second grid's holes are $0.6~\textrm{mm}$, designed to optimize extraction efficiency. Finally, the RIT-2.5 is equipped with a strategically located gas inlet on the periphery, providing enhanced functionality near the extraction zone. This design not only maximizes performance but also showcases advanced engineering principles in plasma technology.

\section{\label{ExptSetup}Experimental setup }

\begin{figure}[!h]
\centering
\includegraphics[width=80mm]{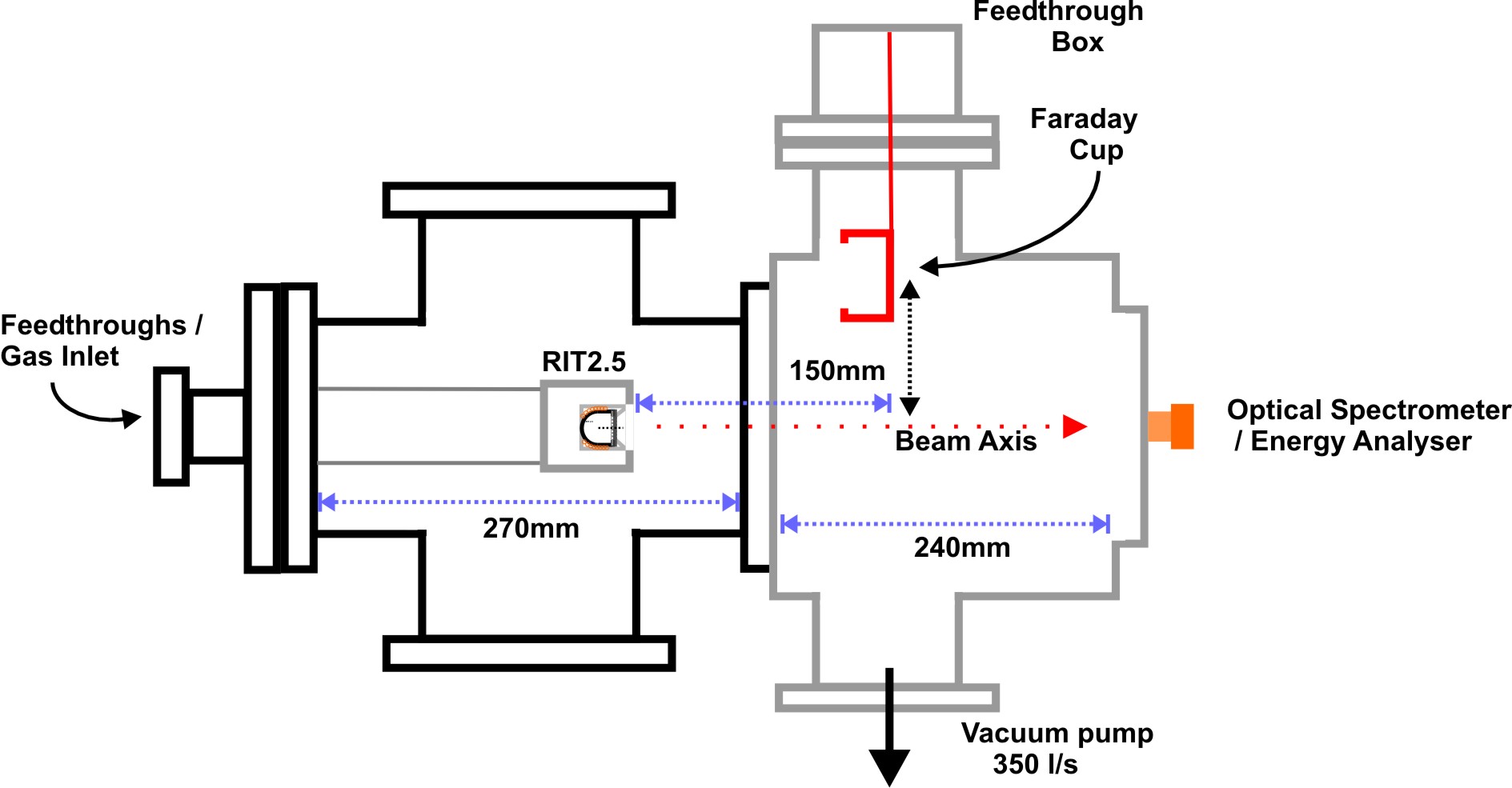}
\caption{\label{fig:expt_setup} Schematics showing experimental setup with ion thruster mounted inside a vacuum chamber and a tank fitted with diagnostics equipments.}
\end{figure}

The experiments were carried out using a dedicated vacuum facility designed to characterize the performance of the RIT-2.5 radiofrequency ion thruster. A schematic of the setup is shown in Fig.~\ref{fig:expt_setup}. The thruster was mounted inside a cylindrical stainless-steel vacuum chamber (diameter: $150~mm$, length: $270~mm$), which was evacuated by a turbomolecular pump backed by a dry scroll pump, reaching a base pressure of $<5\times10^{-6}$~mbar. During operation with xenon propellant, the working pressure was typically in the range of $(2–5)\times10^{-5}$~mbar, depending on the mass flow rate.

The discharge chamber of the RIT-2.5 was powered by a radiofrequency (RF) generator operated at a fixed driving frequency of $2.2$~MHz. The forward power was adjusted by varying the applied RF voltage between $14$ and $20$~V in increments of $2$~V, corresponding to an input power range of $22$–$46$~W. Xenon mass flow rates were controlled using a calibrated mass flow controller, and varied between $0.12$ and $0.24$~sccm. These operating conditions were chosen to cover the typical working regime of the RIT-2.5 while ensuring reproducibility across measurement campaigns.

Downstream of the thruster, a diagnostics chamber was coupled to the vacuum tank to host a suite of non-intrusive and intrusive plasma diagnostics. The extracted ion current was measured using a Faraday cup placed on the thruster axis at a distance of $L=150$~mm. The ion energy distribution was characterized using a retarding field energy analyzer (RFEA) positioned at the same axial location. Additionally, optical emission spectroscopy (OES) was employed to estimate the electron temperature inside the discharge chamber. For this purpose, light emitted from the plasma through an optical viewport was collected using a lens system and directed into a fiber-coupled Czerny–Turner spectrometer (resolution $\Delta \lambda \approx 0.1$~nm).

This configuration enabled simultaneous monitoring of plasma generation within the discharge chamber and characterization of the extracted ion beam, thereby providing a consistent dataset for comparison with numerical simulations.

\section{\label{E-Temp}Electron Temperature Determination from Optical Emission Spectroscopy}

It is possible to determine the electron temperature of a plasma from measurements of electron–atom interactions. In this work, optical emission spectroscopy (OES) was employed for this purpose, using the line-ratio method to extract the electron temperature from selected xenon emission lines.

A full and accurate description of the plasma radiation processes would require a collisional–radiative (CR) model. However, if two transitions originate from levels that are populated and decay primarily through direct electron impact excitation and spontaneous emission, the simpler line-ratio technique described by J. Boffard can be applied\cite{Boffard}. 
In the framework of the coronal model, the photon flux of a spectral line arising from a transition $i\rightarrow j$ is given by

\begin{equation}
\Phi = n_e n_0 \int\limits_0^\infty Q_{(ij)}(E) f(E) \left(\frac{2E}{m_e}\right)^{1/2} dE
\end{equation}

where $n_e$ is the electron density, $n_0$ is the neutral atom density, $Q_{(ij)}(E)$ is the electron impact excitation cross-section, and $f(E)$ is the electron energy distribution function (EEDF) \cite{McWhirter_Corona}. 
For a Maxwellian EEDF,

\begin{equation}
f(E, T_e) = \frac{2\sqrt{E}}{\sqrt{\pi}(kT_e)^{3/2}} \exp\left( -\frac{E}{kT_e} \right) ,
\end{equation}

where $T_e$ is the electron temperature.

In practice, the ratio of photon fluxes from two spectral lines is often used, since it eliminates the dependence on electron density and depends only on $T_e$ :

\begin{equation}
\frac{\Phi_{ij}^{Obs}}{\Phi_{ab}^{Obs}} = \frac{\int\limits_{E_1}^\infty Q_{ij}(E) \exp\left( - E/kT_e \right) E ~ dE}{\int\limits_{E_2}^\infty Q_{ab}(E) \exp\left( - E/kT_e \right) E ~dE} ,
\end{equation}

where $E_1$ and $E_2$ are the excitation threshold energies of the respective transitions. In this study, two emission lines corresponding to different excited states were selected to determine $T_e$.

\begin{figure}[!h]
\centering
\includegraphics[width=82mm]{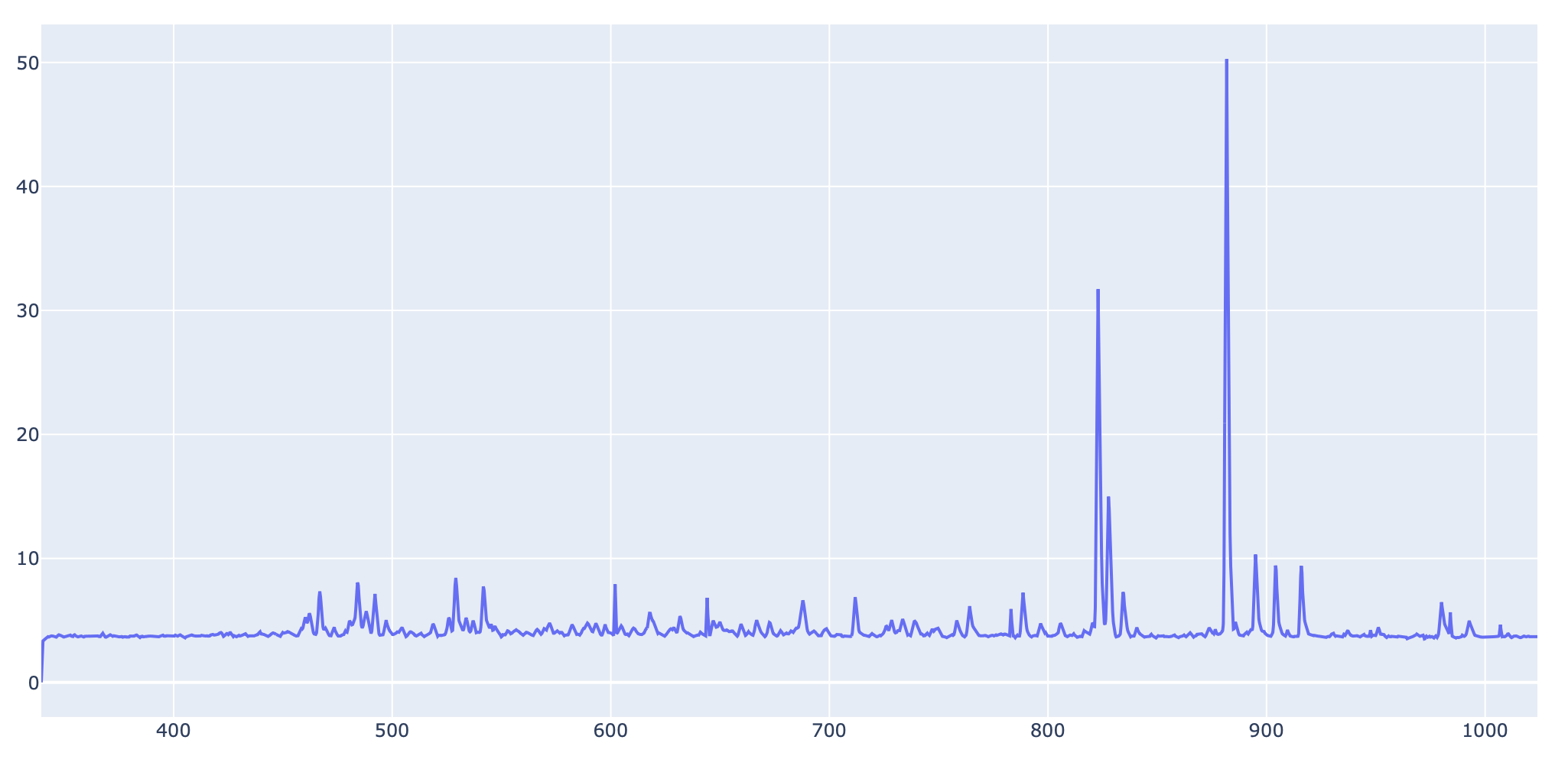}
\includegraphics[width=82mm]{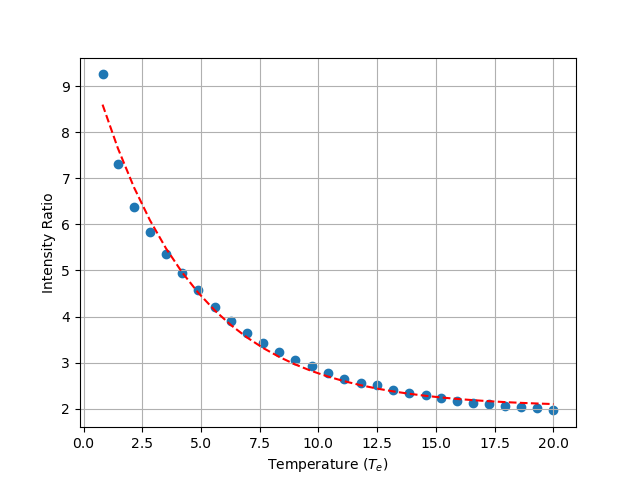}
\caption{\label{fig:expt_measure6} Representative optical emission spectrum of xenon plasma, indicating the selected diagnostic lines.}
\end{figure}

The CR model implemented in the LPP0D code \cite{CRModel} was used to compute the dependence of electron temperature on the measured line intensity ratio. This model was previously validated in a master’s thesis at Laboratoire de Physique des Plasmas, École Polytechnique, where it was shown that the 823–828 nm line pair provided good agreement with experimental results. Fig.\ref{fig:expt_measure6} shows a representative optical spectrum obtained with 38 W RF power deposition and a neutral xenon mass flow rate of 0.22 sccm.
The selected xenon lines at 823 and 828 nm were chosen for their distinct excitation thresholds (11.8 and 12.2 eV) and minimal cascade contributions; the CR model accounts for direct electron impact excitation and radiative decay but neglects stepwise ionization and metastable quenching.

For the examined operating conditions, the plasma electron temperature remains nearly constant in the range of 3–7 eV for a fixed set of input parameters. However, the extracted ion beam current exhibits significant variation with changes in mass flow rate and deposited RF power.

\section{\label{FC}Energy analysis and density estimation using a Retarding Field Energy Analyzer (RFEA)}

The energy distribution of the extracted ion beam was characterized using a retarding field energy analyzer (RFEA). The analyzer employed in this study follows the classical design consisting of two concentric cylindrical electrodes and a downstream collector \cite{SchultePhd2013}. The first electrode was kept at ground potential, while a variable negative bias was applied to the second electrode to establish a retarding potential barrier for ions. Ions with kinetic energy lower than the applied potential were rejected, while higher-energy ions passed through to be collected.

\begin{figure}[h!]
\centering
\includegraphics[width=62mm]{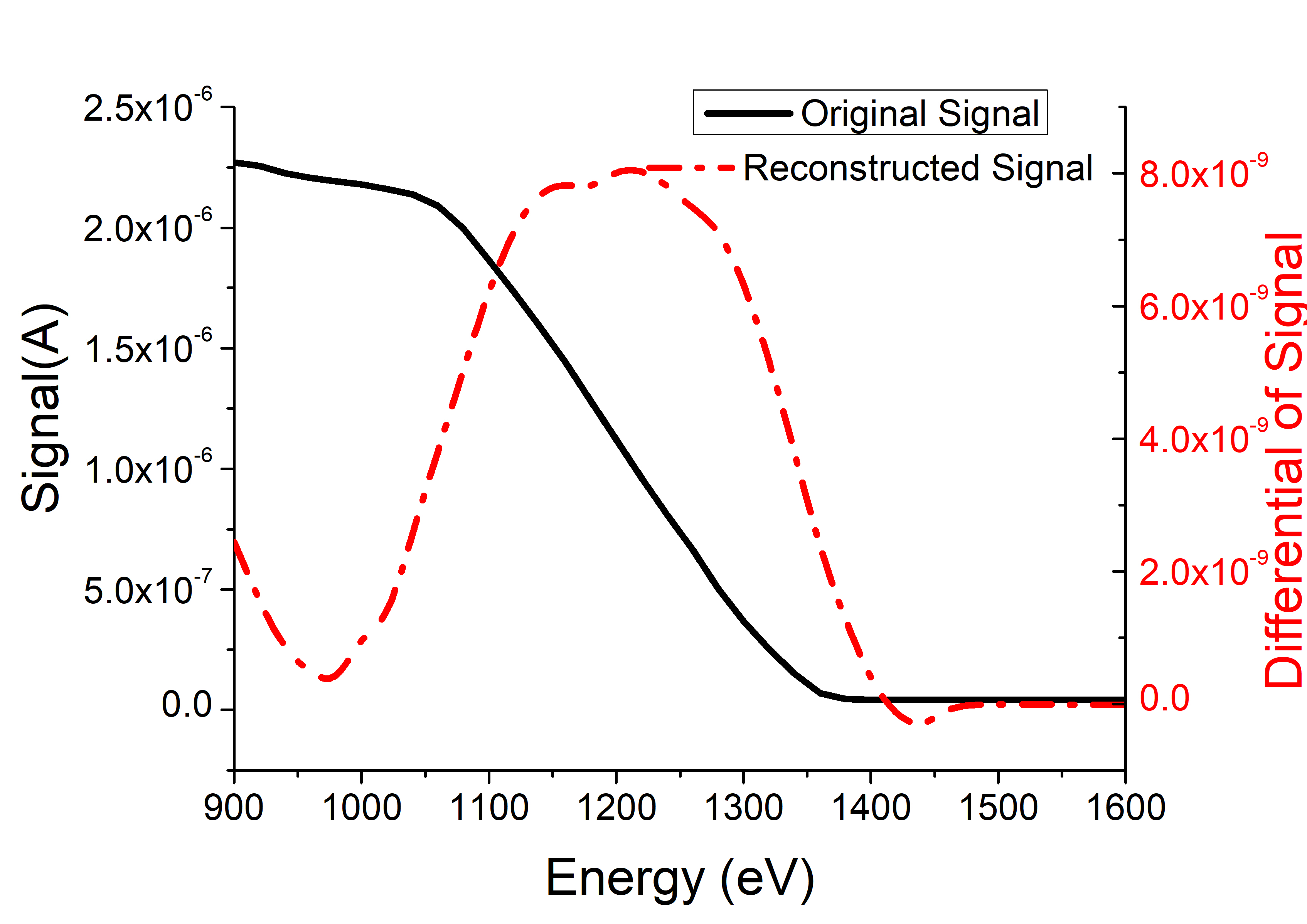}
\caption{\label{fig:expt_measure_energy} 
Measured retarding field energy analyzer (RFEA) signal and the reconstructed energy spectrum of the extracted ion beam at $1300~\mathrm{eV}$ extraction voltage.}
\end{figure}

The collector current, measured as a function of retarding voltage, represents the integrated ion flux with energies above the barrier. The differential energy distribution function (EDF) of the beam can then be obtained from the derivative $dI/dV$, where $I$ is the measured ion current and $V$ the retarding voltage. An example of a measured current-voltage trace and its corresponding reconstructed EDF is shown in Fig.~\ref{fig:expt_measure_energy}. In this particular case, the extraction voltage was set to $1300~\mathrm{V}$, and the EDF exhibits a clear peak corresponding to the accelerated ion population.

The retarding potential was swept in 1 V increments over a 0–1500 V range with a sweep rate of 0.5 V/ms; the resulting I–V curves were smoothed using a Savitzky–Golay filter prior to numerical differentiation to obtain the EDF. Estimated uncertainty in energy resolution was ±3 eV.

\subsection{\label{Sim}Comparison with simulations}

\begin{figure}[h!]
\centering
\includegraphics[width=82mm]{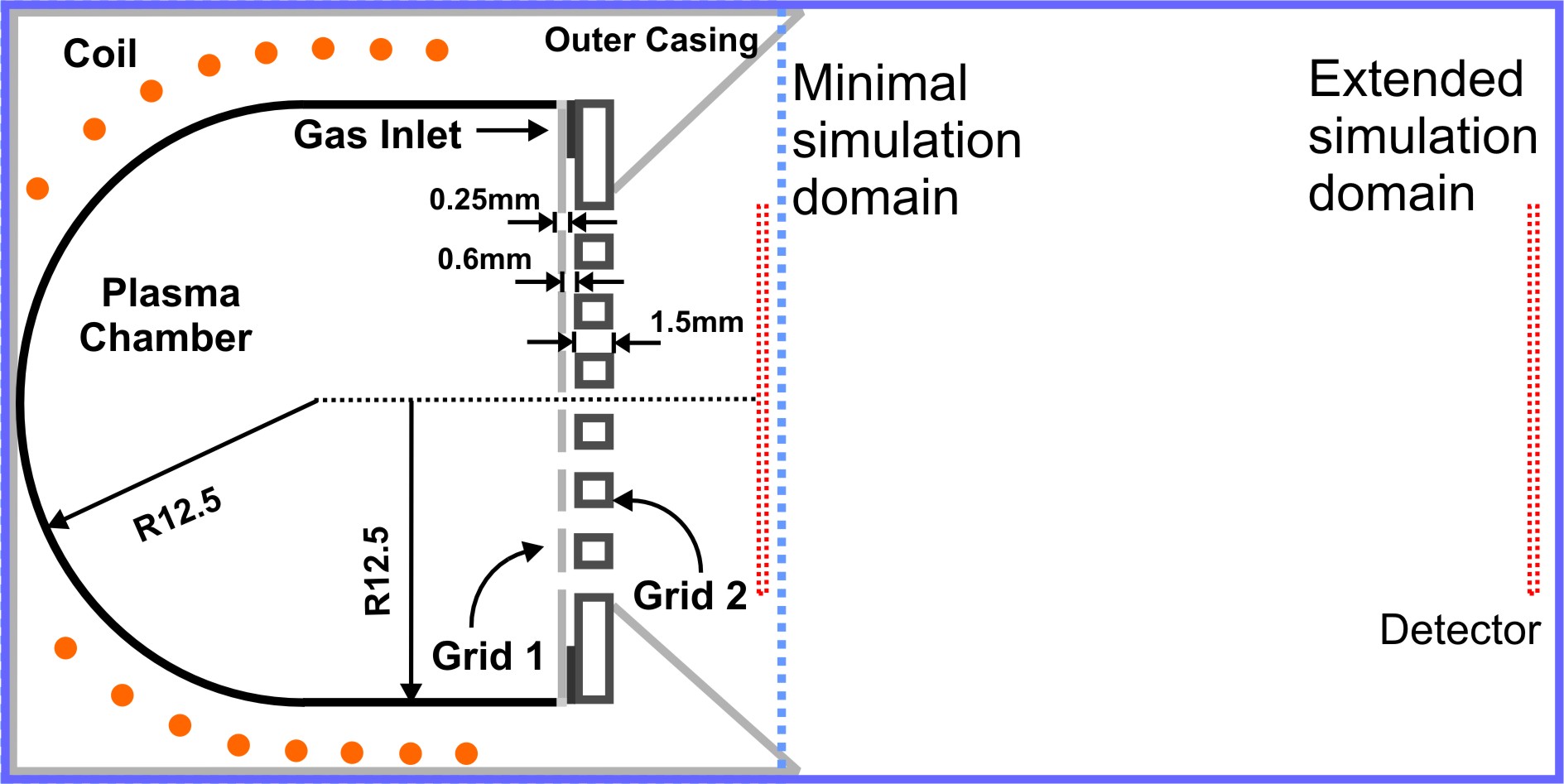}
\caption{\label{fig:sim} 
Two-dimensional projection of the simulation domain. The minimal domain was used for plasma characterization, while the extended domain enabled modelling of ion extraction and beam diagnostics. A planar detector was defined downstream to record energy and phase-space distributions.}
\end{figure}

To interpret the measured EDFs, particle-in-cell (PIC) simulations were performed using the PlasmaPIC code \cite{Joshi2024,Joshi2023,HenrichPhd2013}. The simulation domain is illustrated in Fig.~\ref{fig:sim}, comprising both a reduced minimal domain (for discharge plasma characterization) and an extended domain (for ion extraction studies). A virtual detector plane of $1~\mathrm{mm}$ thickness was placed upstream of the computational boundary to record ion trajectories and velocities. This allowed reconstruction of longitudinal energy distributions and transverse phase-space diagrams of the extracted ions.

\begin{figure}[h!]
\centering
\includegraphics[width=60mm]{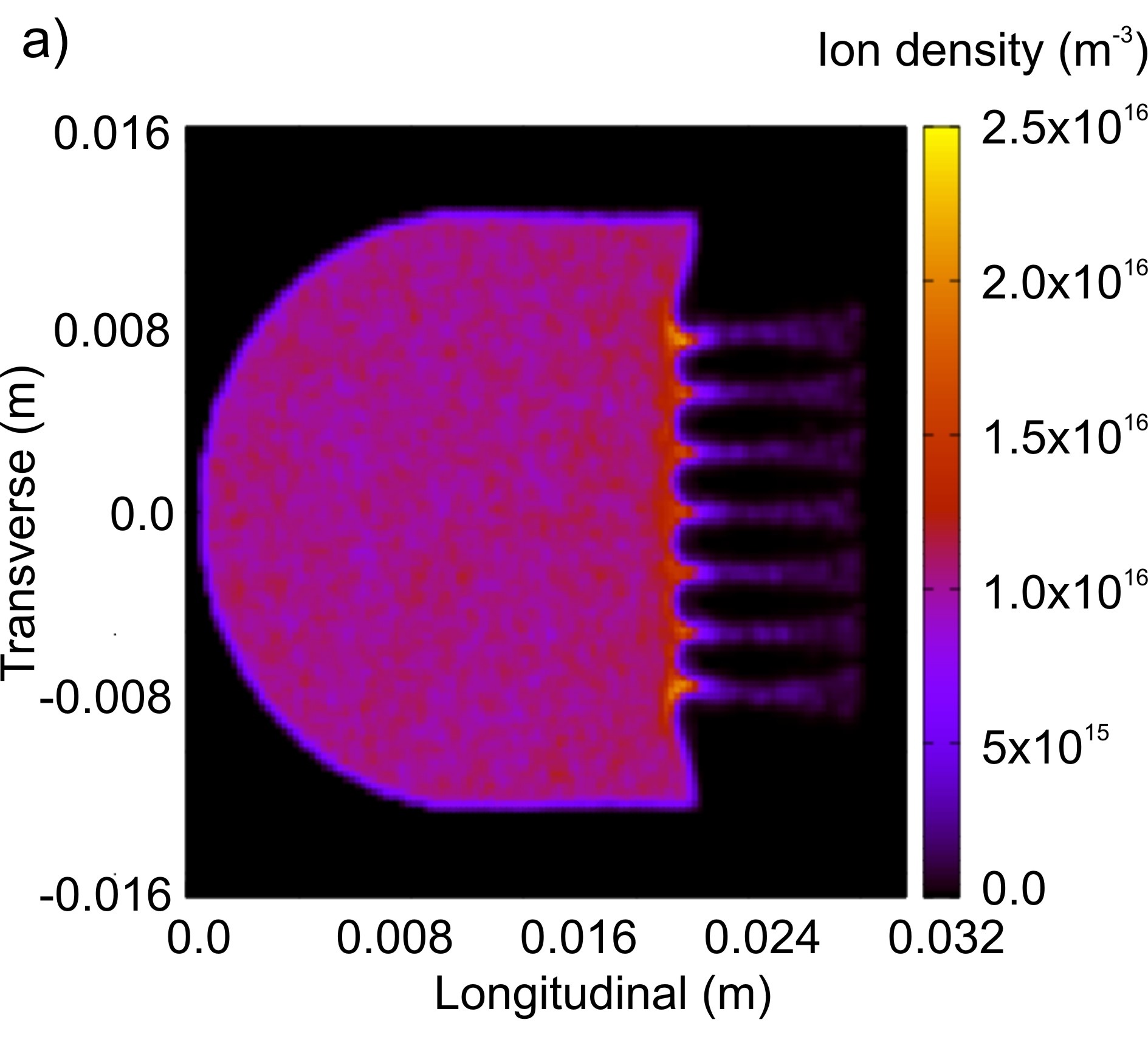}
\includegraphics[width=50mm]{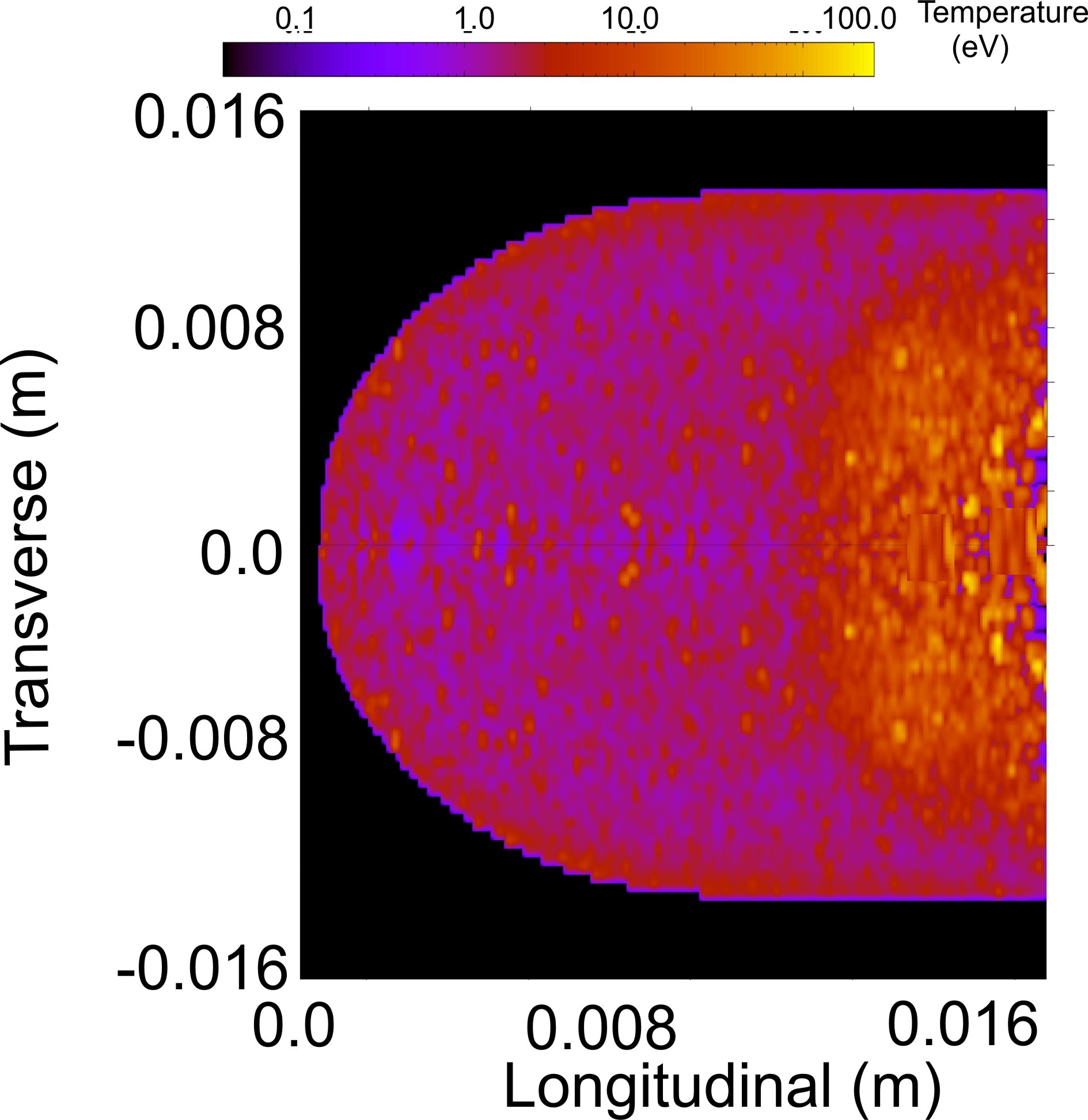} \\
\includegraphics[width=60mm]{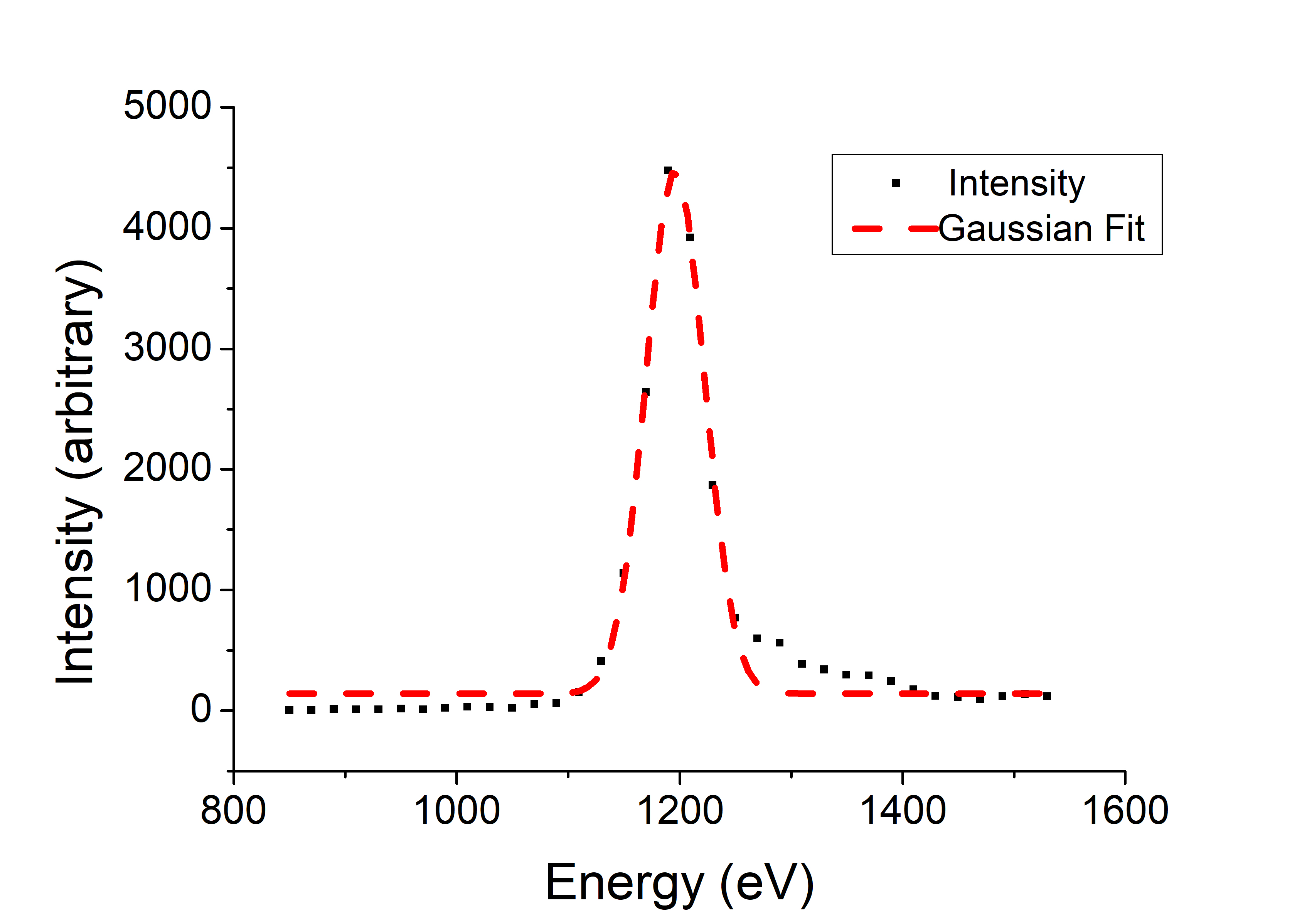}
\includegraphics[width=50mm]{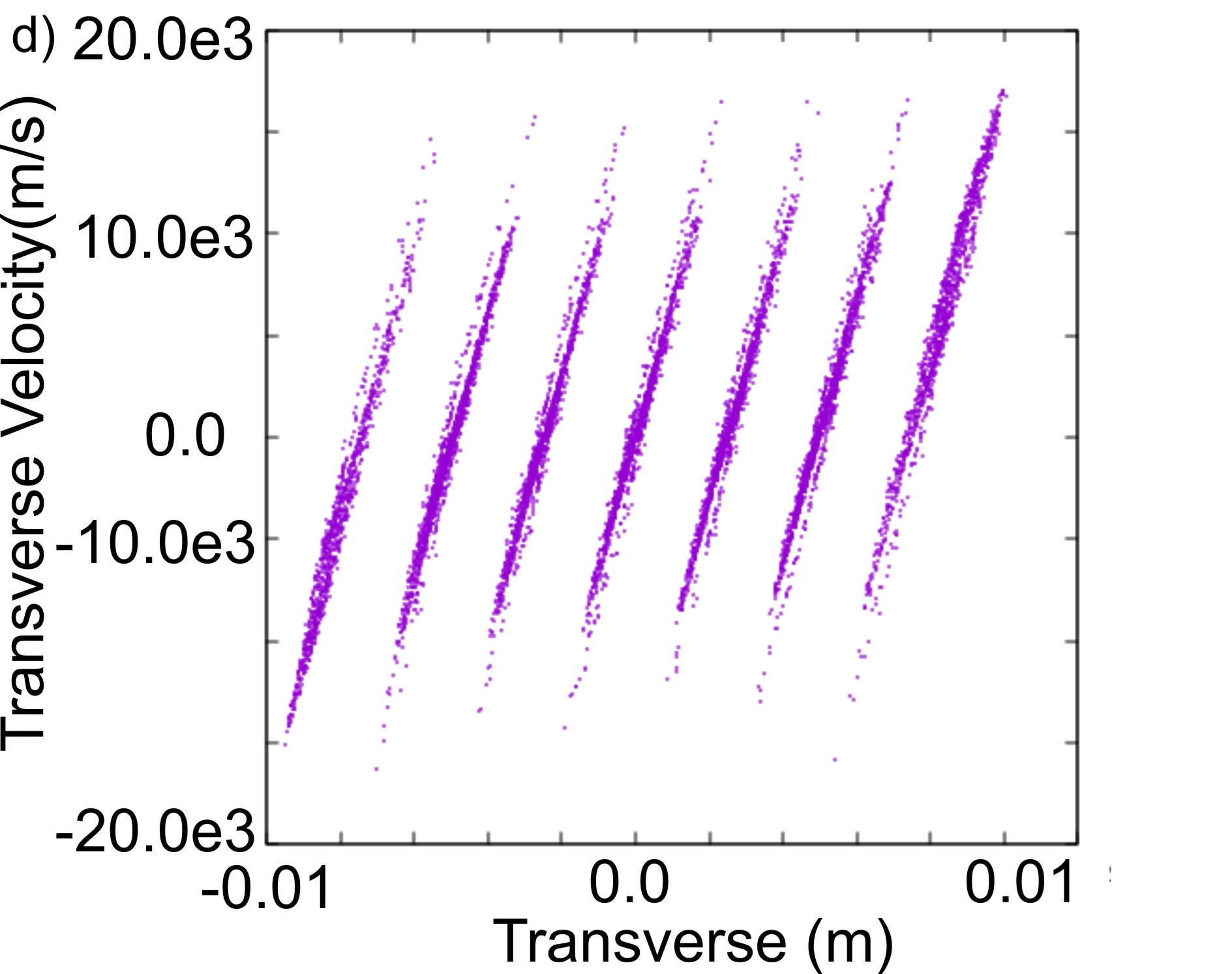}
\caption{\label{fig:sim_results} 
Simulation results obtained using PlasmaPIC. (a) Ion density distribution, (b) electron temperature distribution, (c) energy spectrum just after the extraction grid, and (d) downstream phase-space distribution.}
\end{figure}

Representative results are shown in Fig.~\ref{fig:sim_results}. Figures~\ref{fig:sim_results}a-b illustrate the simulated ion density and electron temperature distributions within the plasma chamber. A plasma sheath is observed near the extraction apertures, with accelerated ions forming a well-defined beam. The extracted ion current was recorded and the ion phase space was reconstructed. Figure~\ref{fig:sim_results}c displays the energy distribution at a detector plane located close to the grid, where a narrow peak with minimal spread is observed. At farther downstream locations (Figure~\ref{fig:sim_results}d), the transverse phase-space distribution reflects the multi-aperture geometry of the grid system.

In this example, a plasma with density $n_e \sim 10^{14}~\mathrm{m^{-3}}$ and electron temperature $T_e = 2~\mathrm{eV}$ was simulated. From these results, average plasma parameters were extracted and compared against the experimental EDFs.  

The PIC simulations employed absorbing boundary conditions at the walls, a Maxwellian electron population with fixed $T_e$. 
The code was previously benchmarked against 1D sheath models and a capacitively coupled plasma discharge

\subsection{\label{E-Den}Plasma density estimation from EDFs}

\begin{figure}[h!]
\centering
\includegraphics[width=80mm]{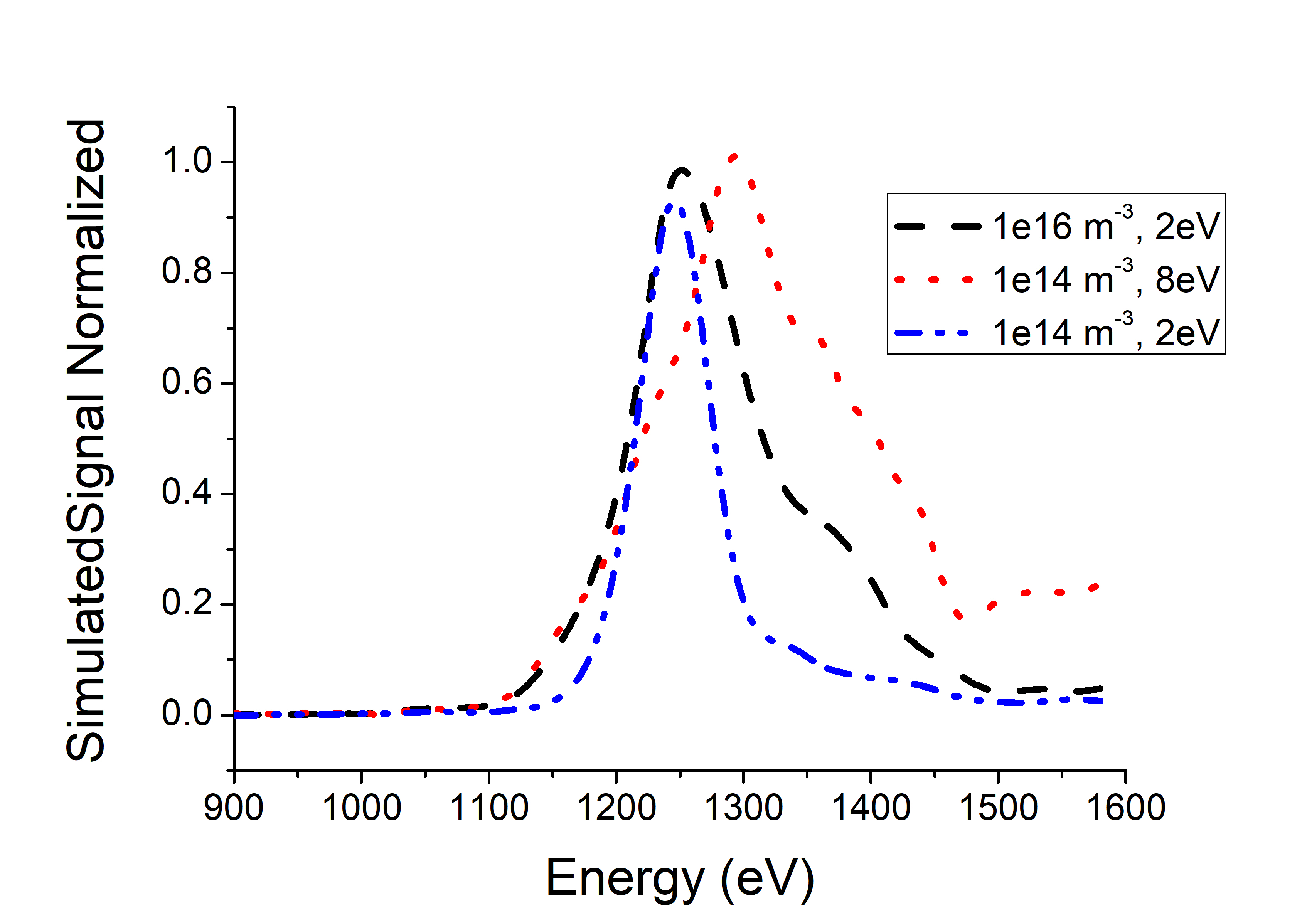}
\caption{\label{fig:expt_measure_sim} 
Simulated ion energy spectra for different plasma densities and electron temperatures, showing the dependence of peak position and energy spread.}
\end{figure}

Direct density measurements inside the discharge chamber could not be performed due to the lack of intrusive diagnostics (e.g., Langmuir probes). Instead, EDF broadening measured with the RFEA was correlated with plasma density through comparison with simulations. Fig.~\ref{fig:expt_measure_sim} presents simulated EDFs for different densities and electron temperatures. Both the peak position and the spectral width were found to depend on $n_e$ and $T_e$, with higher densities producing broader distributions and slight peak shifts.

\begin{figure}[h!]
\centering
\includegraphics[width=62mm]{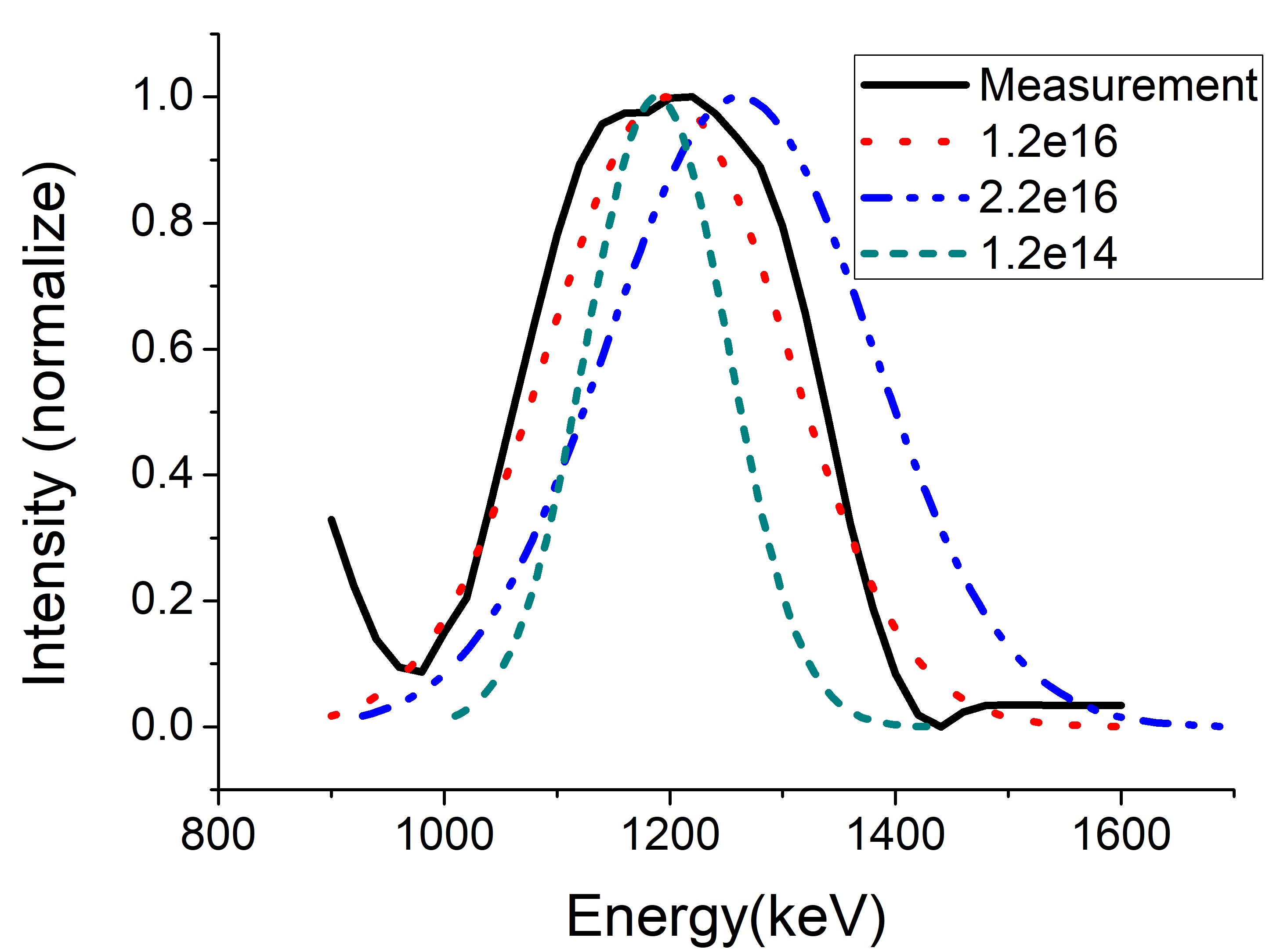}
\caption{\label{fig:expt_measure_energy_2} 
Comparison between measured EDF and simulated spectra for different plasma densities at an extraction voltage of $1200~\mathrm{V}$. The best match corresponds to $n_e \approx 1.2\times10^{16}~\mathrm{m^{-3}}$.}
\end{figure}

Fig.~\ref{fig:expt_measure_energy_2} compares the experimentally measured EDF with simulated distributions. The experimental plasma temperature, estimated from optical emission spectroscopy ($T_e \approx 4.5~\mathrm{eV}$), was used as an input to the simulation. While low-density cases reproduced the peak position but underestimated the width, high-density cases exhibited excessive broadening and peak shifts. The best agreement with experiment was obtained for a plasma density of approximately $1.2\times10^{16}~\mathrm{m^{-3}}$. This correlation enables indirect estimation of plasma density inside the discharge chamber under different operating conditions.

In summary, the combined use of RFEA diagnostics and kinetic simulations provides a robust method for estimating plasma density in regimes where direct probe measurements are not feasible. The approach also highlights the sensitivity of EDF broadening to both density and temperature, which can serve as valuable validation benchmarks for numerical thruster models.

\begin{table*}[htbp]
\centering
\caption{Summary of parameter sweep: simulated plasma densities ($n_e$), electron temperatures ($T_e$), and corresponding full-width at half-maximum (FWHM) of the extracted ion energy distributions (EDF). The cases highlighted in bold correspond to best agreement with experiment.}
\label{tab:paramsweep}
\begin{tabular}{cccccc}
\hline
\textbf{Case} & $n_e~(\times 10^{16}~\mathrm{m^{-3}})$ & $T_e$ (eV) & Extraction Voltage (V) & EDF Peak (eV) & EDF FWHM (eV) \\
\hline
A & 0.5 & 3.0 & 1200 & 1192 & 18 \\
B & 0.8 & 3.0 & 1200 & 1191 & 24 \\
C & 1.0 & 4.0 & 1200 & 1188 & 36 \\
\textbf{D} & \textbf{1.2} & \textbf{4.5} & \textbf{1200} & \textbf{1185} & \textbf{42} \\
E & 1.4 & 5.0 & 1200 & 1182 & 53 \\
F & 1.6 & 6.0 & 1200 & 1178 & 66 \\
\hline
\end{tabular}
\end{table*}

\begin{figure}[htbp]
\centering
\includegraphics[width=82mm]{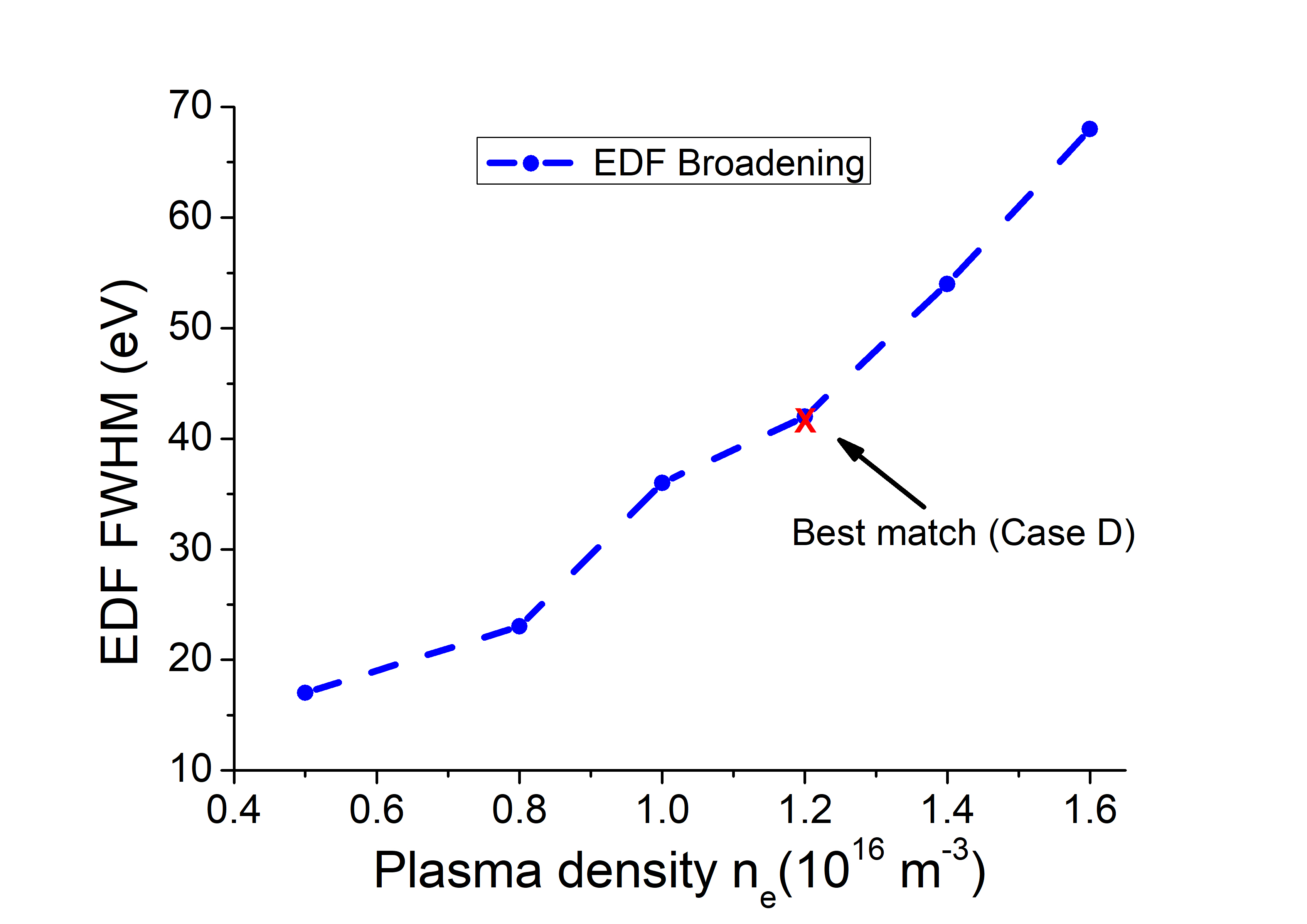}
\caption{Dependence of the extracted ion energy distribution (EDF) full-width at half-maximum (FWHM) on plasma density from the parameter sweep. The red point marks the case (D) that best matches the experimentally measured EDF.}
\label{fig:edfwidth_vs_density}
\end{figure}

A parameter sweep was carried out to investigate the sensitivity of the extracted ion energy distributions (EDFs) to the upstream plasma conditions. Table \ref{tab:paramsweep} summarizes the tested combinations of plasma density ($n_e$) and electron temperature ($T_e$), along with the corresponding EDF peak positions and full-width at half-maximum (FWHM). The results reveal a clear broadening of the EDF with increasing $n_e$ and $T_e$. The trend is illustrated in Fig. \ref{fig:edfwidth_vs_density}, where the FWHM grows approximately linearly with $n_e$. The case highlighted in bold (Case D) provides the closest agreement with the experimentally observed EDF width, indicating that this parameter set most accurately represents the plasma conditions during operation.

\section{Discussion}
\label{Discussion}

While the combined diagnostic–simulation methodology presented here provides valuable insight into RIT-2.5 operation, several limitations and sensitivities should be considered.

First, the electron temperature derived from OES relies on a simplified coronal approximation and a CR model with assumptions regarding metastable states, cascading effects, and Maxwellian electron energy distributions. Deviations from these assumptions could introduce systematic uncertainties of order 10–20\%. Spectrometer calibration, line blending, and optical collection geometry also contribute to the error, warranting a future uncertainty quantification study.

Second, the energy distribution measurements obtained with the RFEA are sensitive to step size, sweep rate, and noise filtering. Fine resolution is needed to resolve sub-eV broadening effects, and small misalignments or local plasma potential fluctuations near the grids may influence the measured spectra. Beam divergence and angular spread, which were not directly measured in this study, could further broaden the apparent EDF.

Finally, the methodology could be extended in several directions. Including diagnostics of beam divergence and total extracted current would allow a more complete thruster performance map. Applying laser-induced fluorescence (LIF) would provide independent velocity distribution measurements for validation. Coupling the kinetic simulations with self-consistent neutral gas dynamics or electromagnetic field solvers could reveal additional optimization pathways, for instance in grid design or RF power deposition strategies.

Overall, this study demonstrates that non-intrusive plasma diagnostics, when combined with validated kinetic simulations, can resolve key internal parameters of compact RF ion thrusters. The same approach could be generalized to other propulsion concepts, including larger RITs, Hall-effect thrusters, and even non-space plasma sources where intrusive probes are impractical.

\section {Conclusions}
\label {Conclusions}

In this work, the key operational parameters of the RIT-2.5 radio-frequency ion thruster were successfully measured using a combination of plasma diagnostics and numerical modelling. Optical emission spectroscopy provided reliable estimates of electron temperature, while retarding field energy analysis enabled reconstruction of the ion energy distribution. Through kinetic simulations, a clear correlation was established between plasma density and the energy spread of the extracted ion beam, providing an indirect method for density estimation in cases where direct probe diagnostics are not feasible.

The experimental results, together with the simulation framework, constitute a comprehensive dataset for validating and refining numerical models of plasma generation and ion extraction in radio-frequency ion thrusters. Moreover, the insights gained from the density–energy spread relationship can be directly applied to optimize thruster performance by guiding adjustments in mass flow, RF power deposition, and extraction conditions. The combined diagnostic and modelling approach presented here thus not only advances the understanding of RIT-2.5 operation but also provides a foundation for further performance optimization, numerical code validation, and future thruster development.

\section*{Data Availability Statement}
Raw data were generated at the  large scale facility HPC Core Facility and the HRZ of the Justus-Liebig-University Giessen. 
The derived data and the experimental data supporting the findings of this study are available from the corresponding author
upon reasonable request.
\appendix

\section{Photographs}

\begin{figure}[htbp]
\centering
\includegraphics[width=42mm]{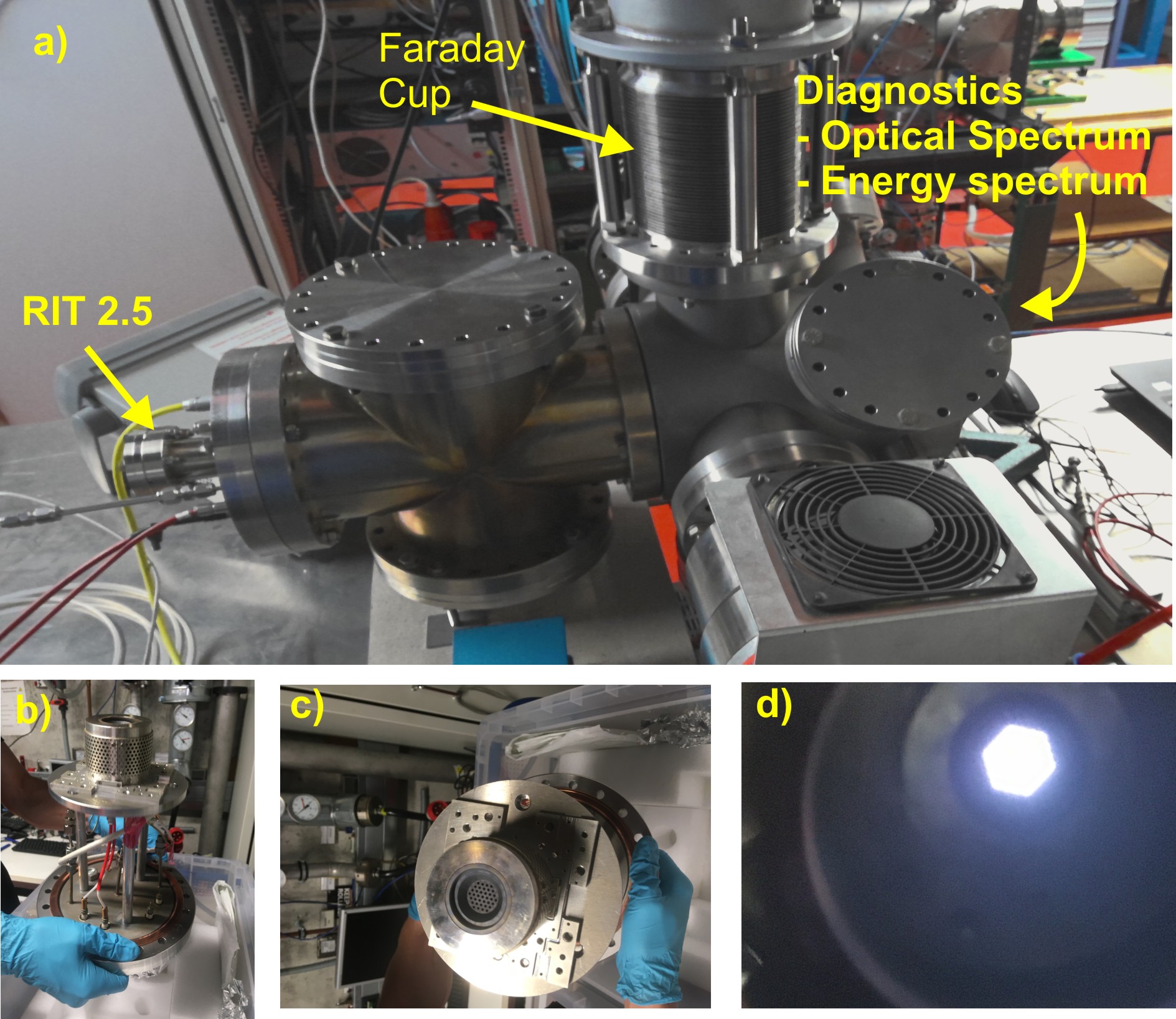}
\caption{Photographs of a ) the experimental setup, b) mounting and feed throughs of the RIT-2.5, c) the extraction aperture arrangement, and the plasma glow of the RIT-2.5.}
\label{fig:photo}
\end{figure}

\nocite{*}
\bibliography{aipsamp}

\providecommand{\noopsort}[1]{}\providecommand{\singleletter}[1]{#1}%
\begin{thebibliography}{19}%
\makeatletter
\providecommand \@ifxundefined [1]{%
 \@ifx{#1\undefined}
}%
\providecommand \@ifnum [1]{%
 \ifnum #1\expandafter \@firstoftwo
 \else \expandafter \@secondoftwo
 \fi
}%
\providecommand \@ifx [1]{%
 \ifx #1\expandafter \@firstoftwo
 \else \expandafter \@secondoftwo
 \fi
}%
\providecommand \natexlab [1]{#1}%
\providecommand \enquote  [1]{``#1''}%
\providecommand \bibnamefont  [1]{#1}%
\providecommand \bibfnamefont [1]{#1}%
\providecommand \citenamefont [1]{#1}%
\providecommand \href@noop [0]{\@secondoftwo}%
\providecommand \href [0]{\begingroup \@sanitize@url \@href}%
\providecommand \@href[1]{\@@startlink{#1}\@@href}%
\providecommand \@@href[1]{\endgroup#1\@@endlink}%
\providecommand \@sanitize@url [0]{\catcode `\\12\catcode `\$12\catcode
  `\&12\catcode `\#12\catcode `\^12\catcode `\_12\catcode `\%12\relax}%
\providecommand \@@startlink[1]{}%
\providecommand \@@endlink[0]{}%
\providecommand \url  [0]{\begingroup\@sanitize@url \@url }%
\providecommand \@url [1]{\endgroup\@href {#1}{\urlprefix }}%
\providecommand \urlprefix  [0]{URL }%
\providecommand \Eprint [0]{\href }%
\providecommand \doibase [0]{http://dx.doi.org/}%
\providecommand \selectlanguage [0]{\@gobble}%
\providecommand \bibinfo  [0]{\@secondoftwo}%
\providecommand \bibfield  [0]{\@secondoftwo}%
\providecommand \translation [1]{[#1]}%
\providecommand \BibitemOpen [0]{}%
\providecommand \bibitemStop [0]{}%
\providecommand \bibitemNoStop [0]{.\EOS\space}%
\providecommand \EOS [0]{\spacefactor3000\relax}%
\providecommand \BibitemShut  [1]{\csname bibitem#1\endcsname}%
\let\auto@bib@innerbib\@empty
\bibitem [{\citenamefont {Kaufman}\ and\ \citenamefont
  {Reader}(1960)}]{Kaufman1960}%
  \BibitemOpen
  \bibfield  {author} {\bibinfo {author} {\bibfnamefont {H.~R.}\ \bibnamefont
  {Kaufman}}\ and\ \bibinfo {author} {\bibfnamefont {P.~D.}\ \bibnamefont
  {Reader}},\ }\href@noop {} {\bibfield  {journal} {\bibinfo  {journal} {Amer.
  Rocket Soc.}\ } (\bibinfo {year} {1960})}\BibitemShut {NoStop}%
\bibitem [{\citenamefont {Loeb}(2002)}]{Loeb2002}%
  \BibitemOpen
  \bibfield  {author} {\bibinfo {author} {\bibfnamefont {H.}~\bibnamefont
  {Loeb}},\ }\bibfield  {title} {\enquote {\bibinfo {title} {Recent work on
  radio frequency ion thrusters},}\ }\href@noop {} {\bibfield  {journal}
  {\bibinfo  {journal} {J. Spacecraft and Rockets}\ } (\bibinfo {year}
  {2002})}\BibitemShut {NoStop}%
\bibitem [{\citenamefont {Loeb}\ \emph {et~al.}(2004)\citenamefont {Loeb},
  \citenamefont {Schartner}, \citenamefont {Weis}, \citenamefont {Feili},\ and\
  \citenamefont {Meyer}}]{Loeb}%
  \BibitemOpen
  \bibfield  {author} {\bibinfo {author} {\bibfnamefont {H.}~\bibnamefont
  {Loeb}}, \bibinfo {author} {\bibfnamefont {K.~H.}\ \bibnamefont {Schartner}},
  \bibinfo {author} {\bibfnamefont {S.}~\bibnamefont {Weis}}, \bibinfo {author}
  {\bibfnamefont {D.}~\bibnamefont {Feili}}, \ and\ \bibinfo {author}
  {\bibfnamefont {B.}~\bibnamefont {Meyer}},\ }\bibfield  {title} {\enquote
  {\bibinfo {title} {Development of rit- microthrusters},}\ }\href@noop {}
  {\bibfield  {journal} {\bibinfo  {journal} {proceedings of 55th International
  Astronautical Congress}\ } (\bibinfo {year} {2004})}\BibitemShut {NoStop}%
\bibitem [{\citenamefont {Samples}\ and\ \citenamefont
  {Wirz}(2019)}]{Samples2019}%
  \BibitemOpen
  \bibfield  {author} {\bibinfo {author} {\bibfnamefont {S.~A.}\ \bibnamefont
  {Samples}}\ and\ \bibinfo {author} {\bibfnamefont {R.~E.}\ \bibnamefont
  {Wirz}},\ }\bibfield  {title} {\enquote {\bibinfo {title} {Development status
  of the miniature xenon ion thruster},}\ }\href@noop {} {\bibfield  {journal}
  {\bibinfo  {journal} {Proceedings of 36th International Electric Propulsion
  Conference}\ } (\bibinfo {year} {2019})}\BibitemShut {NoStop}%
\bibitem [{\citenamefont {Yang}\ and\ \citenamefont {et~al}(2022)}]{Yang2022}%
  \BibitemOpen
  \bibfield  {author} {\bibinfo {author} {\bibfnamefont {X.}~\bibnamefont
  {Yang}}\ and\ \bibinfo {author} {\bibnamefont {et~al}},\ }\bibfield  {title}
  {\enquote {\bibinfo {title} {Numerical simulation of the start-up process of
  a miniature ion thruster},}\ }\href@noop {} {\bibfield  {journal} {\bibinfo
  {journal} {Plasma Sci. Technol.}\ }\textbf {\bibinfo {volume} {24}} (\bibinfo
  {year} {2022})}\BibitemShut {NoStop}%
\bibitem [{\citenamefont {Koizumi}\ \emph {et~al.}(2014)\citenamefont
  {Koizumi}, \citenamefont {Komurasaki}, \citenamefont {Aoyama},\ and\
  \citenamefont {Yamaguchi}}]{Koizumi2014}%
  \BibitemOpen
  \bibfield  {author} {\bibinfo {author} {\bibfnamefont {H.}~\bibnamefont
  {Koizumi}}, \bibinfo {author} {\bibfnamefont {K.}~\bibnamefont {Komurasaki}},
  \bibinfo {author} {\bibfnamefont {J.}~\bibnamefont {Aoyama}}, \ and\ \bibinfo
  {author} {\bibfnamefont {K.}~\bibnamefont {Yamaguchi}},\ }\bibfield  {title}
  {\enquote {\bibinfo {title} {Engineering model of the miniature ion
  propulsion system for the nano-satellite: Hodoyoshi-4},}\ }\href@noop {}
  {\bibfield  {journal} {\bibinfo  {journal} {Trans. JSASS Aerospace Tech}\ ,\
  \bibinfo {pages} {19--24}} (\bibinfo {year} {2014})}\BibitemShut {NoStop}%
\bibitem [{\citenamefont {Okawa}\ and\ \citenamefont
  {et~al}(2004)}]{Okawa2004}%
  \BibitemOpen
  \bibfield  {author} {\bibinfo {author} {\bibfnamefont {Y.}~\bibnamefont
  {Okawa}}\ and\ \bibinfo {author} {\bibnamefont {et~al}},\ }\href@noop {}
  {\bibfield  {journal} {\bibinfo  {journal} {Proc. of 40th AIAA/ASME/SAE/ASEE
  Joint Propulsion Conference and Exhibit}\ } (\bibinfo {year}
  {2004})}\BibitemShut {NoStop}%
\bibitem [{\citenamefont {Ott}\ and\ \citenamefont {Penningsfeld}(1992)}]{Ott}%
  \BibitemOpen
  \bibfield  {author} {\bibinfo {author} {\bibfnamefont {W.}~\bibnamefont
  {Ott}}\ and\ \bibinfo {author} {\bibfnamefont {F.}~\bibnamefont
  {Penningsfeld}},\ }\bibfield  {title} {\enquote {\bibinfo {title} {Beam
  divergence and ion current in multiaperture ion sources},}\ }\href@noop {}
  {\bibfield  {journal} {\bibinfo  {journal} {Max-Planck-Institut fuer
  Plasmaphysik, Garching (Germany)}\ } (\bibinfo {year} {1992})}\BibitemShut
  {NoStop}%
\bibitem [{\citenamefont {Boffard}, \citenamefont {Lin},\ and\ \citenamefont
  {DeJoseph}(2004)}]{Boffard}%
  \BibitemOpen
  \bibfield  {author} {\bibinfo {author} {\bibfnamefont {J.~B.}\ \bibnamefont
  {Boffard}}, \bibinfo {author} {\bibfnamefont {C.~C.}\ \bibnamefont {Lin}}, \
  and\ \bibinfo {author} {\bibfnamefont {C.~A.~J.}\ \bibnamefont {DeJoseph}},\
  }\bibfield  {title} {\enquote {\bibinfo {title} {Application of excitation
  cross sections to optical plasma diagnostics},}\ }\href@noop {} {\bibfield
  {journal} {\bibinfo  {journal} {Journal of Physics D: Applied Physics}\
  }\textbf {\bibinfo {volume} {37}} (\bibinfo {year} {2004})}\BibitemShut
  {NoStop}%
\bibitem [{\citenamefont {McWhirter}(1965)}]{McWhirter_Corona}%
  \BibitemOpen
  \bibfield  {author} {\bibinfo {author} {\bibfnamefont {R.~W.~P.}\
  \bibnamefont {McWhirter}},\ }\enquote {\bibinfo {title} {Plasma diagnostic
  techniques},}\ \ (\bibinfo  {publisher} {Academic Press},\ \bibinfo {year}
  {\noopsort{}1965})\ pp.\ \bibinfo {pages} {201--261}\BibitemShut {NoStop}%
\bibitem [{\citenamefont {Slimane}(2020)}]{CRModel}%
  \BibitemOpen
  \bibfield  {author} {\bibinfo {author} {\bibfnamefont {T.~B.}\ \bibnamefont
  {Slimane}},\ }\emph {\bibinfo {title} {A Xenon Collisional Radiative Model
  for Electric Propulsion Application- Determining the electron temperature in
  a Hall-effect Thruster}},\ \href@noop {} {\bibinfo {type} {M.{S}. thesis}},\
  \bibinfo  {school} {New York University}, \bibinfo {address} {LPP, Ecole
  Polytechnique} (\bibinfo {year} {2020})\BibitemShut {NoStop}%
\bibitem [{\citenamefont {Schulte}(2013)}]{SchultePhd2013}%
  \BibitemOpen
  \bibfield  {author} {\bibinfo {author} {\bibfnamefont {K.}~\bibnamefont
  {Schulte}},\ }\emph {\bibinfo {title} {Studies on the Focusing Performance of
  a Gabor Lens Depending on Nonneutral Plasma Properties}},\ \href@noop {}
  {\bibinfo {type} {{Ph.D.} thesis}},\ \bibinfo  {school} {Goethe
  Universität}, \bibinfo {address} {Frankfurt am Main} (\bibinfo {year}
  {2013})\BibitemShut {NoStop}%
\bibitem [{\citenamefont {Joshi}\ and\ \citenamefont
  {Heiliger}(2024)}]{Joshi2024}%
  \BibitemOpen
  \bibfield  {author} {\bibinfo {author} {\bibfnamefont {N.}~\bibnamefont
  {Joshi}}\ and\ \bibinfo {author} {\bibfnamefont {C.}~\bibnamefont
  {Heiliger}},\ }\bibfield  {title} {\enquote {\bibinfo {title}
  {Particle-in-cell simulation of radio-frequency ion thruster rit-1.0},}\
  }\href@noop {} {\bibfield  {journal} {\bibinfo  {journal} {J Electr Propuls}\
  }\textbf {\bibinfo {volume} {3}} (\bibinfo {year} {2024})}\BibitemShut
  {NoStop}%
\bibitem [{\citenamefont {Joshi}\ and\ \citenamefont
  {Heiliger}(2023)}]{Joshi2023}%
  \BibitemOpen
  \bibfield  {author} {\bibinfo {author} {\bibfnamefont {N.}~\bibnamefont
  {Joshi}}\ and\ \bibinfo {author} {\bibfnamefont {C.}~\bibnamefont
  {Heiliger}},\ }\bibfield  {title} {\enquote {\bibinfo {title}
  {Fluid-kinetic-hybrid simulation for ion thruster using polymorphic
  particles},}\ }\href@noop {} {\bibfield  {journal} {\bibinfo  {journal} {J
  Electr Propuls}\ }\textbf {\bibinfo {volume} {2}} (\bibinfo {year}
  {2023})}\BibitemShut {NoStop}%
\bibitem [{\citenamefont {Henrich}(2013)}]{HenrichPhd2013}%
  \BibitemOpen
  \bibfield  {author} {\bibinfo {author} {\bibfnamefont {R.}~\bibnamefont
  {Henrich}},\ }\emph {\bibinfo {title} {Development of a plasma simulation
  tool for radio frequency ion thrusters}},\ \href@noop {} {\bibinfo {type}
  {{Ph.D.} thesis}},\ \bibinfo  {school} {Justus-Liebig-Universität}, \bibinfo
  {address} {Giessen} (\bibinfo {year} {2013})\BibitemShut {NoStop}%
\bibitem [{\citenamefont {Holste}(2020)}]{Holste}%
  \BibitemOpen
  \bibfield  {author} {\bibinfo {author} {\bibfnamefont {K.}~\bibnamefont
  {Holste}},\ }\bibfield  {title} {\enquote {\bibinfo {title} {Ion thrusters
  for electric propulsion: Scientific issues developing a niche technology into
  a game changer},}\ }\href@noop {} {\bibfield  {journal} {\bibinfo  {journal}
  {Rev. Sci. Instrum.}\ }\textbf {\bibinfo {volume} {91}} (\bibinfo {year}
  {2020})}\BibitemShut {NoStop}%
\bibitem [{\citenamefont {Patterson}\ and\ \citenamefont
  {Foster}(1991)}]{Patterson1991}%
  \BibitemOpen
  \bibfield  {author} {\bibinfo {author} {\bibfnamefont {M.~J.}\ \bibnamefont
  {Patterson}}\ and\ \bibinfo {author} {\bibfnamefont {J.~E.}\ \bibnamefont
  {Foster}},\ }\bibfield  {title} {\enquote {\bibinfo {title} {Ion energy
  diagnostics using retarding potential analyzers},}\ }\href@noop {} {\bibfield
   {journal} {\bibinfo  {journal} {Review of Scientific Instruments}\ }\textbf
  {\bibinfo {volume} {62}} (\bibinfo {year} {1991})}\BibitemShut {NoStop}%
\bibitem [{\citenamefont {Chabert}\ and\ \citenamefont
  {Braithwaite}(2011)}]{Chabert2011}%
  \BibitemOpen
  \bibfield  {author} {\bibinfo {author} {\bibfnamefont {P.}~\bibnamefont
  {Chabert}}\ and\ \bibinfo {author} {\bibfnamefont {N.}~\bibnamefont
  {Braithwaite}},\ }\enquote {\bibinfo {title} {Physics of radio-frequency
  plasmas},}\ \ (\bibinfo  {publisher} {Cambridge University Press},\ \bibinfo
  {year} {\noopsort{}2011})\BibitemShut {NoStop}%
\bibitem [{\citenamefont {Goebel}\ and\ \citenamefont
  {Katz}(2008)}]{Goebel2008}%
  \BibitemOpen
  \bibfield  {author} {\bibinfo {author} {\bibfnamefont {D.~M.}\ \bibnamefont
  {Goebel}}\ and\ \bibinfo {author} {\bibfnamefont {I.}~\bibnamefont {Katz}},\
  }\href@noop {} {}\bibinfo {type} {Tech. Rep.}\ \bibinfo {number} {AL944}\
  (\bibinfo  {institution} {JPL Space Science and Technology Series},\ \bibinfo
  {year} {2008})\ \bibinfo {note} {fundamentals of Electric Propulsion: Ion and
  Hall Thrusters}\BibitemShut {NoStop}%
\end{thebibliography}%


\providecommand{\noopsort}[1]{}\providecommand{\singleletter}[1]{#1}%
%

\end{document}